# RC3E: Provision and Management of Reconfigurable Hardware Accelerators in a Cloud Environment

Oliver Knodel and Rainer G. Spallek

Department of Computer Science
Technische Universität Dresden
Dresden, Germany
{firstname.lastname}@tu-dresden.de

*Abstract*—**Heterogeneous systems consisting of general-purpose processors and different types of hardware accelerators are becoming more and more common in HPC systems. Especially FPGAs provide a promising opportunity to improve both performance and energy efficiency of such systems. Adding FPGAs to clouds or data centers allows easy access to such reconfigurable resources. In this paper we present our cloud service models and cloud hypervisor called RC3E, which integrates virtualized FPGA-based hardware accelerators into a cloud environment. With our hardware and software framework, multiple (virtual) user designs can be executed on a single physical FPGA device. We demonstrate the performance of our approach by implementing up to four virtual user cores on a single device and present future perspectives for FPGAs in cloud-based data environments.**

## I. Introduction

Multi-core and multi-threaded processors were in recent years combined with special dedicated hardware accelerators to improve the performance of applications. It is foreseeable that the future of hardware development lies in more and more massively parallel architectures as used in embedded as well as high performance systems [1]. Especially field-programmable gate arrays (FPGAs) provide an energy-efficient way to achieve high performance by tailoring hardware directly to the application.

Numerous application areas requiring high processing capability employ simple computation cores, data structures and algorithms which are highly suitable for the utilization of FPGAs. In particular the use of reconfigurable hardware to accelerate computationally intensive applications has increased steadily over the last decade [2]. FPGAs provide customized hardware performance and low power consumption which makes them interesting for the field of high performance computing and leads to the new discipline of high performance reconfigurable computing (HPRC) [3].

Due to the growing deployment of reconfigurable hardware as accelerators in HPC systems and data centers, it is necessary to simplify the access to such resources and to increase their usability. One possible solution is provided by the integration of reconfigurable hardware in cloud architectures. Cloud computing is a key technology with the potential to transform the whole information technology industry. The idea is in the end that "using 1,000 servers for one hour costs no more than using one server for 1,000 hours" [4]. Providing reconfigurable hardware in such way can raise its acceptance in many scientific and economic fields by accelerating application.

In contrast to conventional computing resources, the integration of reconfigurable hardware into a cloud infrastructure often proves to be difficult and is currently a topic of research only. This article describes our concept for the flexible integration of FPGAs in a multi-user system, making them available to a broader group of users as a cloud service in a data center. For this purpose, the provision in a distributed multi-user environment and a central administration is necessary [5]. We also introduce our concept of virtual FPGAs (vFPGA), which enables one physical FPGA to host multiple vFPGAs from different users simultaneously. Our approach increases the utilization and efficiency even for small user designs.

A resource management system for physical and especially virtual FPGAs cannot work efficiently without an integrated computing framework providing a virtualization environment. Thus, we implement a fully integrated computing framework, allowing easy access to the FPGA resources through common interfaces on hardware and software level and achieving high throughput communications. The virtualization necessary to provide up to four virtual user designs on a single device is another important feature of our approach. Transferring an application or an algorithm to reconfigurable hardware requires fundamental and profound understanding of the hardware. To reduce the development time of computationally intensive applications, an integration of state-of-the-art high-level synthesis (HLS) tools is necessary and also part of the framework.

The following Section II introduces comparable concepts and related research in the field of reconfigurable hardware in cloud architectures and computing frameworks. Section III gives an overview on cloud service models. Section IV shows the implementation of our resource management system with an overview of our FPGA computing framework. As an example, an algorithm is transferred to the FPGA using HLS in Section V. Section VI concludes and gives an outlook.

## II. Related Work

Reconfigurable hardware such as FPGAs can be used in data centers for hardware acceleration of special applications with simple data structures and streams. In these cases the reason for the use of FPGAs lies, in addition to their high processing speed, mainly in the comparatively low energy consumption compared to graphic processing units (GPUs) and common processors (CPUs). For cloud services, it is furthermore possible to use FPGAs for anonymization of user requests [6] and to increase security [7]. The integration of reconfigurable hardware in cloud architectures is shown in [8]. An example for the integration of hardware accelerators in the open source cloud management system OpenStack is shown in [9]. A comparable contribution with stronger focus on the transfer of applications into an FPGA grid for high performance computing is shown in [10]. Both applications focus on a single cloud service model with a background acceleration of applications using FPGAs.

A cloud integration of reconfigurable resources requires the virtualization of the resource FPGA. The VirtualRC [11] uses a uniform hardware/software interface to realize communication on different FPGA platforms. BORPH [12] takes a similar approach with a homogeneous interface for hard- and software. The FPGA paravirtualization pvFPGA [13] uses a profound integration of an FPGA device driver in a Xen virtual machine.

Furthermore, there exists a number of frameworks for FPGA hardware accelerators with different features. These range from simple PCIe implementations which provide memory transfers [14, 15], to complex frameworks that also incorporate the integration of DRAM and allow for dedicated computational cores [16, 17]. The cores can be generated by external high-level synthesis tools. The OpenCPI framework [17] includes hardware-specific modules such as PCIe, Ethernet and DDR memory, in which the user application is embedded as a computing core in a data flow architecture. Also Leap [18] presents an interesting framework with a so-called FPGA operating system as management core and the possibility to transfer an application to the FPGA using HLS.

Another related topic which has been arising in recent years are the so-called remote laboratories. Their idea is it to access FPGAs in a server room from the workplace or even from home. Such systems are mainly used in universities for teaching and research purposes [19, 20]. These concepts offer the opportunity to share lab resources by time multiplexing, and to save lab equipment, space and costs [20]. Such systems are a kind of special FPGA cloud architecture in a university and education environment [21].

In contrast to the systems mentioned, our aim is it to build a system with various cloud service models enabling remote FPGA labs for university education, hardware acceleration for HPC and also background acceleration for data centers with multiple users on the same physical FPGA.

## III. Overview and Cloud Service Models

The main component of our system is the hypervisor RC3E introduced in Section IV. It manages the resources and provides access to the FPGA devices. The hypervisor is complemented by our computing framework RC2F realizing the virtual partitioning of the FPGAs. By this, multiple users can share the same physical device to maximize utilization. Both the framework and the interface between user applications on hard- and software are presented in Section IV-D.

The crucial point of an integration of FPGAs into a cloud architecture is it to define possible application areas and service models. Before we describe our cloud architecture's hard- and software level, we discuss service models for FPGAs in a cloud environment. In the following we introduce three key service perspectives and compare them with the definition of service models in cloud computing [22]. Fig. 1 gives an overview of our three models and their user-modifiable components.

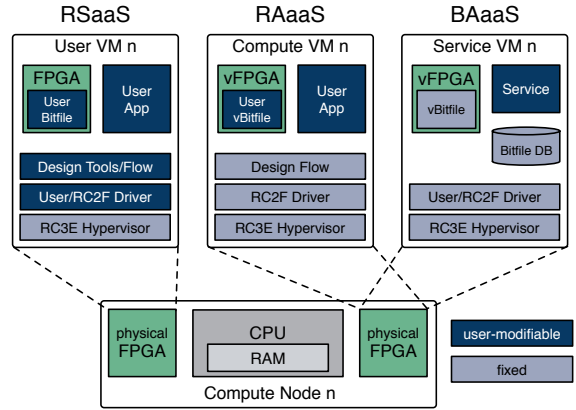

Figure 1: The three service models provided in our cloud environment. In the RSaaS model, users can allocate full physical FPGAs. The RAaaS- and BAaaS model allow multiple concurrent user designs on a single physical FPGA.

### A. Reconfigurable Silicon as a Service – RSaaS

Providing full access to the reconfigurable resource, in this model the user can allocate a complete physical FPGA and can implement the hardware of his choice. Allocation and programming are possible with the management framework provided in Section IV. For hardware interface and driver development fully virtual machines with the necessary FPGA devices attached are allocatable by users.

The allocation of vFPGAs is also possible and increases the utilization and efficiency even for small user designs. In this model the whole development flow is provided as a cloud service. The ability to run multiple design flows simultaneously can greatly reduce design exploration time. Parallel to the software flow, the implementation on real hardware including validation and test can be performed on different FPGAs. Since the model allows users to reconfigure the FPGA, it opens new attack vectors that do not exist in current cloud environments. The concept can be compared to the cloud service models Platform as a Service (PaaS) and Infrastructure as a Service (IaaS).

Application areas include for example education and research. It is possible to access FPGAs in a server room from the workplace or even from home, which offers the opportunity to share lab resources and to save lab equipment, space and costs [19, 21, 23].

### B. Reconfigurable Accelerators as a Service – RAaaS

Another model with less freedom for the user is the Reconfigurable Accelerators as a Service (RAaaS) model, which is inspired by the HPCaaS concept. In this model the FPGA is used as a simple hardware accelerator and the computing framework we introduce in Section IV-D. Only vFPGAs of different sizes are visible, allocatable and useable by the user. The framework provides a communication API on the host as well as FIFO and memory interfaces on the FPGA. The user only has to design the computation core inside the vFPGA and a host program to send and receive data, which reduces development time and optimizes the design process. Such restrictions furthermore have the advantage that the system is significantly safer than the RSaaS model. The RAaaS model can be compared to the PaaS model.

The concept is suitable especially for research- and development-oriented applications as in this field a limitation of hardware resources can be a bottleneck. Our resource management system provides an integrated batch system for long-running applications without direct user interaction. Moreover, a host program can be submitted to a batch system and program the vFPGA region by itself.

### C. Background Acceleration as a Service – BAaaS

Our third model is suitable for applications and services running in common data centers. The vFPGA is not directly visible or accessible by the users. Instead, available applications and services are visible. These services are using vFPGAs in the background to accelerate specific applications. The pre-build bitfiles and host applications are offered by the cloud service provider. Resource allocation and vFPGAs reconfiguration occurs in the background using our resource management system. Because this model provides concrete service applications to the user, it is similar to the Software as a Service (SaaS) model. Application areas include security relevant tasks [6, 7] and in particular computationally intensive routines.

## IV. Reconfigurable Common Cloud Computing Environment

The **R**econfigurable **C**ommon **C**loud **C**omputing **E**nvironment – **RC3E** – is our FPGA hypervisor and grants access through a middleware. The system includes resource management and monitoring of FPGA resources. In the following we introduce the FPGA cloud's hardware architecture in Section IV-A, the hypervisor's software architecture in Section IV-B and Section IV-C. Our FPGA computing framework is presented in Section IV-D and the typical design flow is described in Section IV-E.

### A. Hardware Infrastructure

Our infrastructure consists of nodes with one processor each and up to two physical FPGA boards. The FPGAs are tightly coupled to the processors using PCIe, nodes are connected to each other via Gigabit Ethernet interconnect. To avoid communication bottlenecks, which may be caused by applications requiring communication between FPGAs, the FPGAs can directly access the global interconnect. The basic structure is a modification of the concept we introduced in [5]. Each physical FPGA can host up to four virtual FPGAs. The system is accessible via a service and management node. Our current architecture consists of two nodes using Xilinx ML605 and VC707 development boards.

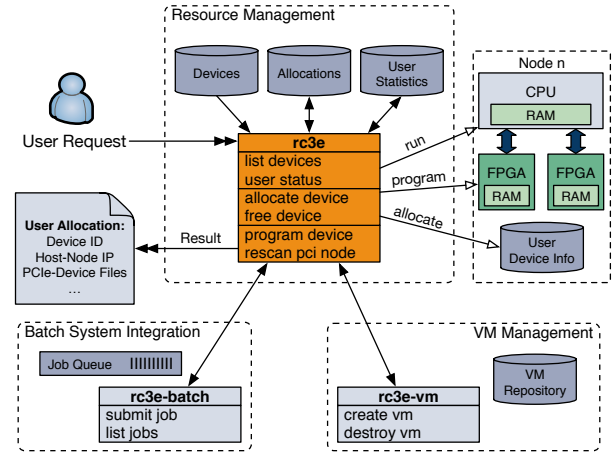

Figure 2: Architecture of the resource management and hypervisor RC3E with additional components (batch system and virtual machines).

### B. Hypervisor

In traditional cloud architectures the hypervisor allows users to run their guest operating systems using virtual CPUs. In our approach the hypervisor allows users to implement and execute their own hardware designs on virtual FPGAs (service models RAaaS and BAaaS in Section III). The RSaaS service model additionally allows for an individual operating system with a physical FPGA.

Due to the existence of multiple nodes, physical FPGAs and also vFPGAs, our RC3E hypervisor acts as a resource manager with load distribution. The overall structure of the RC3E hypervisor is shown in Fig. 2. The hypervisor has access to a database containing all physical and virtual FPGA devices in the cloud system and their allocation status. Each device is assigned to its physical host system (node). If each of the three service models are offered in the cloud simultaneously, all FPGA devices have to be assigned to the RAaaS/BaaS model with vFPGAs and the RC2F framework or to the RSaaS model where framework and virtualization are optional.

In the RAaaS/BaaS service model the FPGAs are configured with a basic design containing the RC2F framework, which provides a PCIe endpoint and basic device status information (see Section IV-D). If no vFPGA is allocated and the device is not allocated, most of the clocks in this design are disabled to reduce power consumption. The resource manager always tries to minimize the number of active vFPGAs and to maximize the utilization of physical FPGAs to thereby reduce energy consumption. In the RSaaS service model the user can allocate a complete physical FPGA, which has to be marked separately in the device database (and is therefore excluded from vFPGA allocations).

### C. Middleware

The RC3E hypervisor is running on the management node and can access each FPGA node. Users can access the cloud services directly through a middleware with a command line interface on the management node. A client middleware running on a client machine will be added in a future version. It is possible to run applications with FPGA acceleration and pre-build bitfiles in the background (BAaaS) without direct allocation of the FPGA resource by the user. The RAaaS model often requires direct interaction, which makes direct allocation of vFPGAs or physical FPGAs with the RC2F framework necessary via the middleware. FPGA configuration and the execution of host applications on the node with the allocated FPGA are possible with separate commands. In both cases, vFPGA configuration is realized by partial reconfiguration (PR).

Table I: Latency of local and remote FPGA status calls and bitstream configuration.

| | RC2F Status | Configuration* | PR |
|---|---|---|---|
| Local without RC3E | 11 ms | 28.370 s | 732 ms |
| Local/Remote Node over RC3E | 80 ms | 29.513 s | 912 ms |

\* Configuration using JTAG and USB

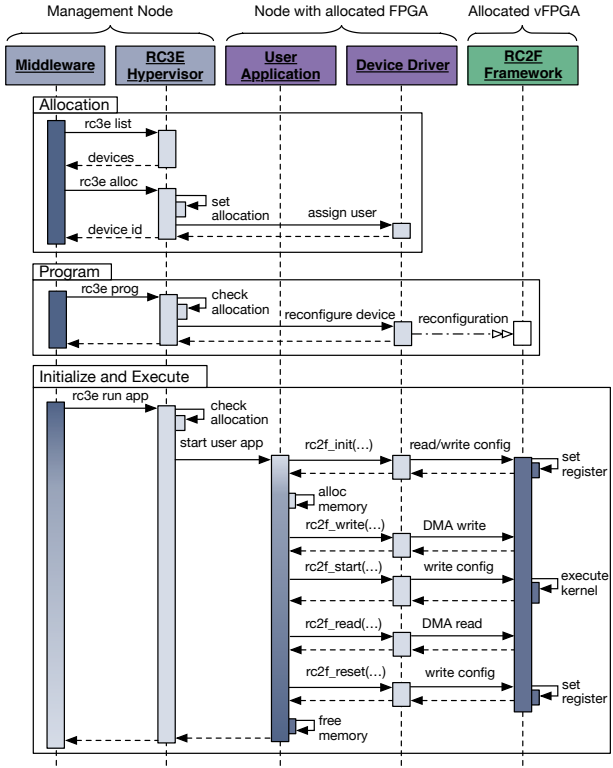

Figure 3: Interaction of middleware, RC3E hypervisor, RC2F user software application and the RC2F framework with the user design on a vFPGA.

The RSaaS model allows full access to the allocated FPGA, but allocation, configuration and execution are performed by the middleware. As the hypervisor implements PCIe hot-plugging by restoration of the PCIe link parameters after reconfiguration, the user can also change the PCIe endpoint on the FPGA. Fig. 3 shows the process of resource allocation, programming, initialization and execution. The overhead caused by the RC3E framework is shown in Table I.

In a typical cloud environment there are always sufficient hardware resources to meet user demands. As our academic test architecture consists of only two nodes with four FPGAs, we integrated a batch system for long-running applications without direct user interaction to improve overall system utilization. A job of the batch system is to specify the type as well as a configuration file for the FPGAs. Furthermore, we integrated the allocation of user-specific virtual machines with direct access to allocated FPGAs as an extension of the RSaaS service model.

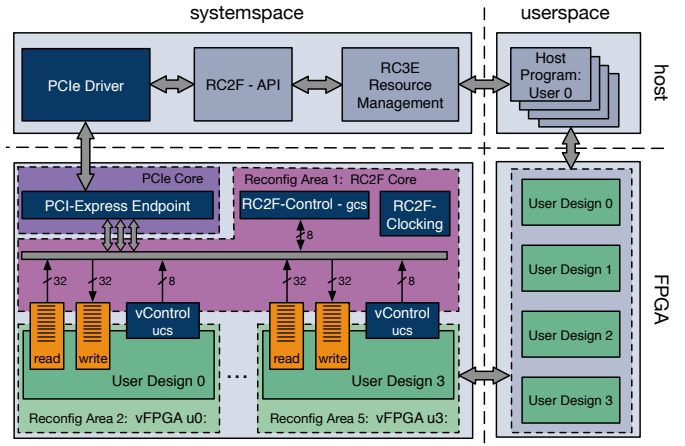

Figure 4: RC2F design with partial reconfiguration areas integrated into a host system.

### D. Computing Framework – RC2F

Hardware acceleration using FPGAs in a cloud environment requires, in addition to resource management and user administration, a framework realizing the vFPGA concept and allowing integration of user cores. We therefore provide the **R**econfigurable **C**loud **C**omputing **F**ramework – **RC2F** – which is fully integrated into our RC3E environment and provides high communication throughput using PCIe. The framework is typically used in our RAaaS and BAaaS models.

*1) Hardware Design:* In the RSaaS service model the user has the freedom to implement the hardware of his choice. Such kind of model opens new attack vectors which can cause physical or functional damage to the system. Thus, writing full bitstreams should only be allowed in research (and educational) systems. The preferred basic design for the RAaaS and BAaaS model in our cloud is provided by our RC2F framework and is shown in the lower left part of Fig. 4. The design can provide up to four vFPGAs with independent user designs.

The main part of the RC2F framework consists of a controller managing the configuration and the user cores as well as the monitoring of status information. The controller's memory space is accessible from the host through the API and on the FPGA via dedicated control signals (full reset, user reset, test loopback, etc.). In- and output-FIFO for streaming applications providing high throughput and memory interfaces for configuration are provided as user interfaces. Table II shows the components' resource utilization of the components for implementations with up to four vFPGAs. On a Xilinx Virtex 7 XC7VX485T the resource utilization for a basic design providing four vFPGAs is less than 3%.

Table II: Resource utilization of the individual components for up to four vFPGAs, throughput and memory latency.

| Component | LUT | FF | BRAM | Latency | Throughput Core (max) |
|---|---|---|---|---|---|
| PCI Endpoint | 3,268 | 3,592 | 8 | | |
| RC2F Control (gcs) | 125 | 255 | 1 | 0.198 ms | |
| 1 vFPGA | 3,689 | 3,127 | 4 | | |
| Total | 7,082 | 6,974 | 13 | 0.208 ms | ≈ 798 MB/s |
| Utilization*(%) | 2.3 | 1.2 | 1.3 | | |
| 2 vFPGAs | 4,414 | 3,790 | 8 | | |
| Total | 7,807 | 7,637 | 17 | 0.221 ms | ≈ 397 MB/s |
| Utilization*(%) | 2.6 | 1.3 | 1.7 | | |
| 4 vFPGAs | 5,139 | 4,471 | 16 | | |
| Total | 8,532 | 8,318 | 25 | 0.273 ms | ≈ 196 MB/s |
| Utilization*(%) | 2.8 | 1.4 | 2.3 | | |

* Xilinx VC707 evaluation board with a XC7VX485T

*2) Communication Interface and Host API:* The main communication between FPGA and host is implemented using PCIe. The low-level implementation is based on an IPCore providing simple device files on the host and FIFO as well as memory interfaces on the FPGA [24]. The throughput of the core is limited to 800 MB/s and will thus be replaced in further versions. Fig. 4 gives an overview of the system divided in host/FPGA and user-/systemspace. The figure also shows the FPGA design with PCIe, RC2F core with global configuration space (gcs) and user cores. As interface to the user cores, a user configuration space (ucs) for user-definable commands is implemented as dual port memory. Streaming access is implemented using asynchronous FIFOs, which also divide the system clock from the user clock. Both components together serve as interfaces between the partial reconfiguration areas. The latency for a configuration memory access (gcs in the RC2F module and ucs in the vFPGAs) and the maximal throughput of the FIFOs for concurrent data transfers are shown in Table II.

On the host the FPGA is accessible by PCIe drivers which provide separate device files for each FIFO and each memory. The RC2F host API interacts with the RC3E hypervisor and provides access to the user-allocated resources without a direct user interaction with the device files. For security reasons the device files are protected by access rights. Because of this additional virtualization layer concurrent users can interact with their allocated devices without influencing each other.

The API calls are inspired by the interaction between host and GPU in the NVIDIA CUDA programming environment [25] or the OpenCL [26] framework. The three basic types are (a) global device control, status query and configuration, (b) user kernel control, status query and reconfiguration and (c) data transfers. Due to security reasons only the RSaaS service model allows such interactions.

### E. Design Flow

The service models using the RC2F framework are also inspired by the CUDA design flow. Separating computation into hardware and software components, the hardware components will be implemented on the FPGA using Xilinx Vivado HLS and will interact with the software components on the host through our API. The entire design flow is shown in Fig. 5. In addition to the RC2F host-library, a HLS library is necessary to provide an integration of the user HLS design into the RC2F hardware and the vFPGAs. HDL components are required for the hardware user interface in the FPGA design. Further extensions of the system will include debugging and tracing of user designs on physical FPGAs.

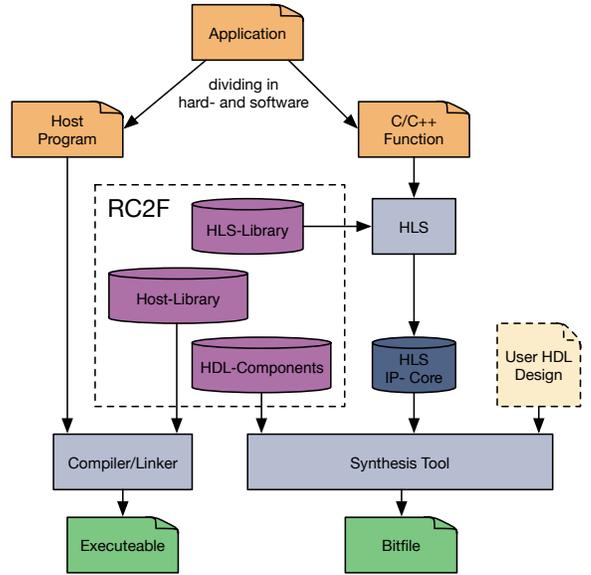

Figure 5: Design flow starting from an application consisting of host program and C-Function as input for HLS.

## V. Example Application and Performance Results

In this section an example application is transferred to a vFPGA using HLS and the presented framework. As application we choose a matrix multiplication which offers both high amounts of data and computational complexity. Moreover, we convert the application to work with a streaming-optimized interface inside our RC2F framework. To reach high throughput we stream the data necessary for 100,000 matrix multiplications through the core. Table III gives an overview of the resources necessary for a $16 \times 16$ matrix multiplication with up to four user cores and a $32 \times 32$ matrix multiplication with up to two user cores.

The host application starts individual parallel user threads sending matrices to the cores, measures runtime and calculates the throughput. For a $16 \times 16$ multiplication the throughput of a single core is compute limited by about 509 MByte/s. Two cores on the same physical FPGA share the bandwidth of 800 MByte/s, which results in a communication bottleneck with a throughput of 398 MByte/s per core and in almost the same runtime as a single core. Four simultaneous user cores affect each other significantly stronger, but in the end the overall performance and the utilization of the physical FPGA is much more efficient.

## VI. Conclusions and Outlook

This paper presents a way to integrate FPGAs as a resource into a cloud environment and to make them available to multiple users. Three possible cloud service models are introduced and integrated into our hypervisor RC3E, allowing resource management for virtual FPGA resources using predefined regions on real devices. The resource management is expanded in a subsequent step by a batch system and by the ability to allocate virtual machines.

Table III: Matrix multiplication Streaming performance (32 Bit Float) with up to four cores (100,000 multiplications each).

| Design | Area | | | | Runtime per Core | Throughput per Core |
|--------|------|------|------|------|------------------|---------------------|
| | LUT | FF | DSP | BRAM | | |
| 16 × 16 | | | | | | |
| 1 vCore | 25,298 | 41,654 | 80 | 14 | 0.73s | 509 MByte/s |
| 2 vCores | 44,408 | 76,963 | 160 | 19 | 0.86s | 398 MByte/s |
| 4 vCores | 81,761 | 146,974 | 320 | 28 | 1.41s | 198 MByte/s |
| 32 × 32 | | | | | | |
| 1 cCore | 64,711 | 125,715 | 160 | 14 | 3.27s | 279 MByte/s |
| 2 vCores | 123,249 | 245,103 | 320 | 19 | 3.43s | 277 MByte/s |

Furthermore, a framework is presented which aims at simplifying the development of computationally intensive applications on FPGAs. In contrast to other approaches the framework is fully integrated into our hypervisor, which significantly increases virtualization possibilities. On a Xilinx Virtex 7 XC7VX485T the resource utilization for a basic design providing four vFPGAs is less than 3%. The paper concludes with an example application using our framework and showing the performance tradeoff between flexibility of virtualized FPGA resources and a dedicated system.

One of the most important next steps is it to introduce a profound security concept for the system. At the moment the RC3E hypervisor works with access protection and only authorized users can program their allocated device. In the future we plan to implement sanity checking for (partial) bitfiles to avoid both damage by a tampered bitstream and access to the parts not reconfigurable by the users as for example physical ports. Such security aspects are essential for the integration of FPGAs into a productive cloud environment.

In future we plan to improve the hypervisor, the security of the system and we will provide debugging opportunities. Furthermore, we will implement a more profound virtualization of the FPGA devices. Currently, both information about the FPGA type and a free predefined vFPGA region are necessary for design flow. We will try to hide these information from the user and to manipulate the partial configuration file to utilize every feasible vFPGA region. A migration of user designs between vFPGAs and physical FPGAs is also intended.